\documentstyle[prl,aps,multicol,epsf,times]{revtex}

\renewcommand{\i}{{\mathrm{i}}}

\newcommand{\im}{\mathop{\mathrm{Im}}}
\newcommand{\tr}{\mathop{\mathrm{tr}}}

\newcommand{\e}{{\mathrm{e}}}
\renewcommand{\d}{{\mathrm{d}}}

\begin{document}
\draft

\title{Exact c-number Representation of Non-Markovian
Quantum Dissipation}
\author{J\"urgen T. Stockburger%
\thanks{present address: Institut f\"ur Theoretische Physik II,
Universit\"at Stuttgart, Pfaffenwaldring 57, 70550 Stuttgart, Germany}
and  Hermann Grabert}
\address{Fakult\"at f\"ur Physik,
Albert-Ludwigs-Universit\"at,
Hermann-Herder-Str. 3,
79104 Freiburg, Germany}
\date{Received }
\maketitle
\begin{abstract}
The reduced dynamics of a quantum system interacting with a linear
heat bath finds an exact representation in terms of a stochastic
Schr{\"o}dinger equation. All memory effects of the reservoir are transformed
into noise correlations and mean-field friction. The classical limit
of the resulting stochastic dynamics is shown to be a generalized
Langevin equation, and conventional quantum state diffusion is
recovered in the Born--Markov approximation. The non-Markovian exact dynamics,
valid at arbitrary temperature and damping strength, is exemplified
by an application to the dissipative two-state system.
\end{abstract}

\begin{multicols}{2}
\narrowtext Irreversible quantum processes are important in almost every
field of condensed-matter physics and chemistry. In phenomena as
different as diffusion of light particles in solids and light
harvesting in biological systems, the dissipation of energy, the
destruction of phase coherence, and the generation of entropy play a
key role. The question how such processes can be described by the
stochastic propagation of quantum states has recently attracted
increasing attention \cite{dalib92dum92breue95,gisin92,perci98,%
diosi96strun96diosi97diosi98strun99,stock98stock99}.
For {\em classical} systems with linear dissipation, Langevin
equations provide a theoretical (and numerical) tool to accurately
describe the interaction of a system with a complex thermal reservoir
in comparably simple terms of stochastic forces and memory friction.
For open {\em quantum} systems, no such simple and exact approach in
terms of an equation of motion has been established so far. The
dynamical quantity of interest is the reduced density matrix, obtained
by tracing out the reservoir degrees of freedom from the
dynamics. Significant information about correlations between system
and reservoir is lost in this operation, hence no deterministic
differential equation of motion can be derived for the reduced density
matrix without approximations.
Traditional approaches treat these correlations perturbatively,
yielding approximate equations of motion such as Redfield or Master
equations. While these have been used very successfully in the fields of
quantum optics and magnetic resonance, there are many
condensed-matter problems where they are qualitatively wrong. Large
coupling constants and long correlation timescales
both need to be treated non-perturbatively.
This can be accomplished in a formally exact manner by path integrals
for open quantum systems\cite{feynm63calde83,grabe88,weiss99,klein95},
but an analytic evaluation of these functional integrals is usually
restricted to special cases and approximations, e.g., the
semiclassical limit. Furthermore, present exact numerical techniques
such as Quantum Monte Carlo (QMC) methods \cite{egger94,makar94} have
to cope with a sign problem arising from the fact that probability
amplitudes with varying phases rather than probabilities have to be
added in a quantum computation.

Quantum State diffusion (QSD)
\cite{dalib92dum92breue95,gisin92,perci98} has been
established as an alternative theory of quantum dissipation in the
perturbative regime. Through the {\em stochastic}\/ propagation of
{\em pure}\/ quantum states it yields an intuitively appealing picture
of the behavior of individual quantum trajectories in open systems and
permits effective numerical calculations. Generalizing this formalism
to the case of non-perturbative, non-Markovian dynamics holds the
promise to overcome most of the above-mentioned limitations of
currently known techniques. This approach has recently been taken by
Di\'osi, Strunz, and Gisin
\cite{diosi96strun96diosi97diosi98strun99}, but at the price of
incorporating non-Markovian retardation effects in a memory
functional. In practice this means that a general solution of the
resulting equation of motion, even numerically, is almost as elusive
as that of the underlying path integral.

In this Letter we derive and discuss stochastic Schr\"o\-ding\-er
equations for open systems which allow the treatment of linear
dissipation of arbitrary strength and correlation time scales. Without
approximation, all effects of the system-reservoir interaction are
re-cast in the form of c-number noise and friction forces with a
suggestive physical interpretation: The ensemble of phase-space points
described by a classical Langevin equation is generalized to a
stochastic ensemble of quantum states. Both descriptions are linked by
a correspondence principle for open systems.

\begin{figure}
\epsfxsize=0.8\columnwidth
\centerline{\epsffile{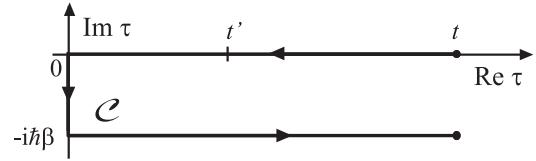}}
\caption[]{Contour for propagators in thermal correlation
functions.}
\label{fig:contour}
\end{figure}

Any general theory of quantum dissipation has to start from an
open system embedded in a large reservoir system, whose degrees of
freedom are treated fully quantum mechanically, but can later be
eliminated from the dynamics. The Hamiltonian of such a model consists
of system and reservoir terms and an interaction potential,
\begin{equation}
H = H_0 + H_R + H_I\;.
\end{equation}
Let us consider a thermal correlation function
\begin{equation}
\langle A B(t) \rangle =
\tr \left\{ \e^{-\beta H} A \e^{{\i\over\hbar} Ht}
B \e^{-{\i\over\hbar} Ht} \right\} / \tr \e^{-\beta H}\;, 
\end{equation}
where $A$ and $B$ are
system operators. The time evolution operators can be joined with the
``imaginary-time'' propagator $\e^{-\beta H}$ to form a propagator
with complex time argument which describes the time evolution of a
quantum system along the contour $\cal C$ depicted in Fig.\
\ref{fig:contour}. This time evolution can be expressed by a path
integral over the configuration space of the system 
\begin{equation}
\langle A  B(t) \rangle \propto
\int {\cal D}[{\bf q}]\, \e^{{\i\over\hbar}S_0[{\bf q}]}
I[{\bf q}] \langle {\bf q}_{0^+} | A | {\bf q}_{0^-}\rangle
\langle {\bf q}_{t^+} | B | {\bf q}_{t^-}\rangle,
\end{equation}
where a normalization factor has been suppressed for simplicity. The
action functional
\begin{equation}
S_0[{\bf q}] =  \int_{\cal C} \d\tau 
L({\bf q},\dot{{\bf q}}) \label{eq:action}
\end{equation}
depends in the usual way on the classical Lagrange function. The
measure $\d\tau$ is the complex time differential on $\cal C$, and
$\dot{{\bf q}} = {\d {\bf q}\over \d\tau}$. In the generalized
influence functional
\begin{equation}
I[{\bf q}] \equiv \left\langle T_{\cal C} \exp \left (
- \int_{\cal C} \d\tau {\i\over\hbar} H_I({\bf q})\right)
\right\rangle_\beta
\end{equation}
the thermal average of the contour-ordered exponential is taken with
respect to the unperturbed reservoir. This exact result can be
simplified \cite{kubo62,weiss99} by performing a cumulant expansion of
the expectation value in $I[{\bf q}]$. The most commonly discussed
case is that of {\em linear dissipation}, equivalent to truncating the
cumulant expansion after the second order. This procedure is exact for
a model reservoir of harmonic oscillators, but also applies to any
microscopic model whose collective response to a perturbation $H_I$ is
dominated by its linear term.

In the following we restrict ourselves to linear dissipation,
for which $I[{\bf q}]$ is a Gaussian functional determined by the
correlation function $\langle H_I({\bf q(\tau)}) H_I({\bf
q}(\tau'))\rangle$.
After an expansion of $H_I({\bf q})$ in terms of an arbitrary set of
reservoir operators $X_j$,
\begin{equation}
H_I = \sum_j f_j({\bf q}) X_j\;,
\end{equation}
the correlation matrix $L_{jk}(\tau-\tau') = \langle X_j(\tau)
X_k(\tau')\rangle_\beta$, characterizing the isolated
reservoir, completely determines the influence functional
\begin{eqnarray}
I[{\bf q}] &=& \exp\Bigg (-{1\over \hbar^2} \sum_{j,k} \int_{\cal C} \d\tau
\int_{\tau' \prec \tau} \d\tau' f_j({\bf q}(\tau))\nonumber\\
&& L_{jk}(\tau-\tau') f_k({\bf q}(\tau')) \Bigg ).
\label{eq:influ}
\end{eqnarray}
Here ``$\prec$'' denotes the order relation induced by the contour
orientation.

The aim of the present work is to transform the exact description of
open quantum systems by the influence functional technique into an
equivalent stochastic propagation of the quantum states of the open
system. We first show that the influence functional
$I[{\bf q}]$ can be constructed as a noise average of the form
\begin{equation}
I[{\bf q}] = \left\langle \exp\Bigg (
{\i\over\hbar} \sum_j \int_{\cal C} \d\tau z_j(\tau)
f_j({\bf q}(\tau)) \Bigg ) \right\rangle_W ,
\label{eq:noisact}
\end{equation}
where the average is taken with a Gaussian probability measure $W[z_j]$.
If the stochastic covariance matrix of the complex-valued noise coordinates
$z_j(\tau)$ matches the quantum correlation matrix of the coupling
operators $X_j$,
\begin{equation}
\langle z_j(\tau) z_k(\tau')\rangle_W = \left\{
\begin{array}{ll}
{1\over 2}L_{jk}(\tau-\tau'), & \tau \succeq \tau' \\
{1\over 2}L_{jk}(\tau'-\tau), & \tau \prec \tau',
\end{array} \right.
\label{eq:zcorr}
\end{equation}
Eqs. (\ref{eq:influ}) and (\ref{eq:noisact}) constitute a formal
identity between Gaussian functionals. Noting that Eq.\
(\ref{eq:zcorr}) does not fully determine the noise statistics, we
find that correlations of the type $\langle z_j(\tau)
z_k^*(\tau')\rangle$ can be chosen such that the covariance matrix of
all real-valued noise components represents a positive quadratic form,
i.e., Gaussian noise $\{z_j\}$ with the desired properties exists.

For each sample of the noise, the propagation is now governed by a
time-local action functional, i.e., the path-integral dynamics can be
translated into an equivalent Schr\"odinger equation. The propagation
along the contour $\cal C$ is governed by the Hamiltonian $H_0$ and a
stochastic potential term,
\begin{equation}
\i \hbar | \dot \psi\rangle = H_0 |\psi\rangle
 - \sum_j z_j(\tau)\, f_j({\bf q}) |\psi\rangle .
\label{eq:single}
\end{equation}
This remarkably simple equation shows that linear dissipation can be
described exactly by a linear QSD theory containing no memory terms.

In order to compare this new finding with previous results, we need to
choose the Feynman-Vernon influence functional\cite{feynm63calde83}, for
which the noise vanishes on the imaginary axis. For simplicity, we
discuss a one-dimensional coupling $H_I = - q x$, where $x$ is a
reservoir operator with correlation function $L(\tau-\tau')$,
associated with a single noise coordinate $z(\tau)$. The propagation
along the two real-time segments of the contour can be re-stated in
the form of a stochastic Liouville equation for the reduced density
matrix,
\begin{equation}
\i\hbar{\d\rho\over\d t} = [H_0,\rho] -z_1 q \rho + z_2^* \rho q,
\label{eq:fveq}
\end{equation}
where $z_1(t) = z(t)$ and $z_2(t) = z(t-\i\hbar\beta)$, and where the
sample $\rho$ of the reduced density matrix is separable, $\rho =
|\psi_1\rangle\langle\psi_2|$. Hence Eq.\ (\ref{eq:fveq}) is just a
compact notation for two stochastic Schr\"odinger equations for
$|\psi_1\rangle $ and $|\psi_2\rangle$.

The noise forces may be represented as the sum of statistically
independent terms, $z_{1,2} = z +
v_{1,2}$. The terms $v_1$ and $v_2$ have identical statistics, but are
uncorrelated. Performing an
average over $v_{1,2}$ takes us back to the partial
decomposition of the influence functional given by Di\'osi, Strunz and
Gisin \cite{diosi96strun96diosi97diosi98strun99}.
Using the same representation, the stochastic Schr\"odinger equation of
Markovian QSD can be recovered.  The same limits and
approximations used to derive Lindblad-type Master equations and their
QSD counterparts from a system-reservoir model, i.e., weak damping and
time coarse-graining, can be applied to
Eq.\ (\ref{eq:fveq}). Coarse-graining is done at an intermediate timescale
which is short compared to the relaxation time of the damped system
but long compared to its oscillation periods and the relaxation times
of the reservoir.  In this approximation the dynamics is only
sensitive to segments of the noise spectrum of $z$ which are centered
around the natural transition frequencies of the undamped system. In
the Born-Markov limit the coarse-graining time scales to zero, and
each of these segments naturally transforms into a distinct Markovian
noise force coupled to a Lindblad operator associated with the
respective quantum transition.
Finally, the work of Stockburger and Mak \cite{stock98stock99} using
real noise and earlier work concentrating on strictly ohmic
damping\cite{KuboKleinert} can be recovered from the present
decomposition of the Feynman-Vernon influence functional by applying
the stochastic decomposition only to the real part of $L(\tau-\tau')$
while explicitly evaluating memory effects induced by the imaginary
part.

As in the linear versions of conventional QSD, $\tr\rho$ is not
conserved by the time evolution of Eqs. (\ref{eq:single}) and
(\ref{eq:fveq}). A transparent physical interpretation of the noise
forces and stochastic samples arises in a modified nonlinear theory
for which the trace of the reduced density matrix is preserved not
only in the stochastic average, but for each sample.

It will be advantageous to introduce new noise variables $\xi = (z_1 +
z_2^*)/2$ and $\hbar\nu = z_1 - z_2^*$, with
resulting covariances
\begin{eqnarray}
\langle \xi(t) \xi(t') \rangle_W &=& {\rm Re} L(t-t')
\label{eq:xiauto}\\
\langle\xi(t)\nu(t')\rangle_W &=& (2\i /\hbar) \Theta(t-t') \im
L(t-t')\nonumber \\
&&= -\i\chi_R(t-t') \label{eq:xinu}\\
\langle\nu(t)\nu(t')\rangle_W &=& 0 , \label{eq:nuauto}
\end{eqnarray}
where $\chi_R$ is the response function of the reservoir.
The noise force $\xi(t)$ can be chosen real, and this case will be
discussed here for simplicity.
 For
comparison with phenomenological friction models, one conventionally
augments the coupling Hamiltonian by a counter-term in order to make
the coupling translationally invariant. The additional term is a quadratic
potential modification $\mu q^2/2$ with $\mu = \int_0^\infty
\d t \chi_R(t)$, which eliminates the static response of the
reservoir. Eq.\ (\ref{eq:fveq}) is thus transformed into
\begin{equation}
\i\hbar\dot\rho = [H_0,\rho] + {\mu\over
 2}[q^2,\rho]
- \xi[q,\rho] -{\hbar\over 2} \nu \{q,\rho\}.
\label{eq:linliou}
\end{equation}
For the normalized density matrix sample $\hat\rho = \rho /
\tr \rho$, this translates into the quasilinear equation of motion
\begin{equation}
\i\hbar\dot{\hat\rho} = [H_0,\hat\rho] + {\mu\over
2}[q^2,\hat\rho] - \xi[q,\hat\rho]
-{\hbar\over 2} \nu \{q-\bar{q},\hat\rho\}
\label{eq:quasilin}
\end{equation}
with $\quad \bar{q} = \tr q\hat\rho$. When averaging the normalized
density matrix $\hat\rho$, the factor
\begin{equation}
\tr \rho = \exp\left(\i \int_0^t \d t'\bar{q}(t')
\nu(t')\right)
\label{eq:measfctr}
\end{equation}
needs to be incorporated into the integration measure.
Because the original integration measure $W[\xi,\nu,\nu^*]$ is
Gaussian, the new measure $W_t[\xi,\nu,\nu^*]$ including the factor
$\tr \rho$ can be rewritten by ``completing the square'' in the
exponent of the Gaussian functional, making the exponent a
quadratic form of shifted variables, and yielding
$W_t[\xi,\nu,\nu^*] = W[\xi_t,\nu_t,\nu^*_t]$ with
\begin{eqnarray}
&& \xi_t(t') = \xi(t') + \int_0^t \!\!\d s\, \chi_R(t'-s)
\bar{q}(s)\\
&& \nu_t(t') = \nu(t')\\
&& \nu^*_t(t') = \nu^*(t') - \i \int_0^t \!\!\d s\,
\langle\nu^*(t')\nu(s)\rangle_W \bar{q}(s)
\end{eqnarray}

Using $W[\xi_t,\nu_t,\nu^*_t]$ as integration measure allows us
to interpret the {\em shifted}\/ variables as noise to be used for
stochastic integration.
The Jacobian determinant of the variable change is unity: Because
$\nu_t=\nu$, and because $\bar{q}$ is independent of $\nu^*$,
it simplifies to
\begin{equation}
\left|\begin{array}{ccc}
{\delta \xi_t\over \delta\xi} & {\delta\xi_t\over\delta\nu} &
{\delta \xi_t\over \delta\nu^*}\\
{\delta \nu_t\over \delta\xi} & {\delta\nu_t\over\delta\nu} &
{\delta \nu_t\over \delta\nu^*}\\
{\delta \nu^*_t\over \delta\xi} & {\delta\nu^*_t\over\delta\nu} &
{\delta \nu^*_t\over \delta\nu^*}
\end{array}\right|
=
\left|\begin{array}{ccc}
{\delta \xi_t\over \delta\xi} & {\delta\xi_t\over\delta\nu} & 0\\
0 & 1 & 0\\
{\delta \nu^*_t\over \delta\xi} & {\delta\nu^*_t\over\delta\nu} & 1
\end{array}\right|
=
\left|\delta\xi_t\over \delta\xi\right|.
\end{equation}
Using the fact that both the bath response and $\bar{q}[\xi,\nu]$ are
causal, one finds that $\left|\delta\xi_t / \delta\xi\right|$ is
represented by a triangular matrix with unit diagonal elements, i.\,e.,
$\left|\delta\xi_t / \delta\xi\right| = 1$.

After the variable change (with subscripts $t$ now dropped)
the equation of motion is rewritten as
\begin{eqnarray}
\i\hbar\dot{\hat\rho} &=& [H_0,\hat\rho] -
 \xi[q,\hat\rho] + m \gamma(t) {\bar q}(0) [q,\hat\rho]\nonumber\\
&&{} + m
\!\int_0^t 
\!\!\d t'\, \gamma(t-t')\dot{\bar{q}}(t')\,[q,\hat\rho] \nonumber\\
&&{} + {\mu\over 2}[(q-\bar{q})^2,\hat\rho]
-{\hbar\over 2} \nu \{q-\bar{q},\hat\rho\}.
\label{eq:eomfriction}
\end{eqnarray}
Here we have integrated by parts, using the
relation $m\dot\gamma(t) = -\chi_R(t)$ between the reservoir response
function and the friction kernel $\gamma(t)$. Remarkably, the mere
choice of a constraint for $\tr\hat\rho$ has taken us from the
abstract mathematical decomposition (\ref{eq:noisact}) to an equation
of motion with a number of terms looking quite familiar---as in
classical dissipation, the dynamics is governed by c-number force
terms representing noise and a mean-field version of classical
friction. Additional terms ascertain that the noise-averaged
$\hat\rho$ contains the effect of all system-reservoir
correlations.

Further examining this analogy, we find a ``sample-by-sample''
correspondence between quantum mechanical and classical stochastic
dynamics. The time evolution of Eq.\ (\ref{eq:eomfriction}) can be
described in a reduced Heisenberg picture, which is defined by
interpreting expectation values of the form $\tr\{A^\dagger
\hat\rho\}$ as a scalar product $(A,\hat\rho)$ of two operators. This
allows the introduction of Heisenberg operators $A(t)$ through the
propagating superoperator ${\cal U}(t)$ and its adjoint,
\begin{equation}
(A,\hat\rho(t)) = (A,{\cal U}(t)\hat\rho_0) =
({\cal U}^\dagger(t)A,\hat\rho_0) = (A(t),\rho_0).
\end{equation}
In the classical limit, terms proportional to $q-\bar{q}$ in
Eq.\ (\ref{eq:eomfriction}) can be neglected, and the equations of
motion governing the evolution of Heisenberg operators $A(t)$ become
those of a classical stochastic system. For a potential model, these
reduce\cite{footnote-parts} to the generalized Langevin equation
\begin{equation}
m\ddot{q}(t) = - V'(q(t)) - m\int_0^t \d t' \gamma(t-t')
\dot{q}(t') + \xi(t)\;.
\end{equation}
The limit $\hbar\to 0$ turns our stochastic ensemble of quantum states
gradually into an ensemble of phase-space points with
classical thermal noise and memory friction, i.e., the quantum
dynamics defined in Eq.\ (\ref{eq:eomfriction}) obeys a stochastic
correspondence principle.

Fig.\ 2 shows numerical results for the symmetric spin-boson
system\cite{footnote-see}, where $q$ is the Pauli matrix $\sigma_z$
and $H_0 = (\hbar\Delta/ 2) \sigma_x$, compared to a QMC algorithm
which directly samples a path integral \cite{egger94}. For the
coherent dynamics shown in Fig.\ 2a, the QMC method does not fully
converge even if run longer than our simulation. Due to an only
marginally ergodic Metropolis random walk, the accuracy of the QMC
method degrades significantly with increasing $t$. Problems of this
kind never appear in our algorithm because our samples are
statistically independent by construction.  The incoherent dynamics
shown in Fig.\ 2b, a forte of the QMC method, is treated equally well
by both approaches.

We have derived exact stochastic equations of motion for the reduced
density matrix in linear form (\ref{eq:single}) and trace-conserving
form (\ref{eq:eomfriction}). In the latter, a close correspondence
with classical Langevin equations is evident. Examples of numerical
results show that the stochastic equations of motion presented here
provide a promising alternative to currently available exact methods
based directly on the path integral representation. Given the freedom
of choice for the noise covariances left undetermined by the physics
of quantum dissipation, future improved numerical methods are to be
anticipated.

The authors would like to thank Walter Strunz for valuable discussions
and correspondence. This work was supported by Deutsche
Forschungsgemeinschaft under SPP 470.


\begin{figure}
\centerline{
\epsfxsize=0.5\columnwidth
\epsffile{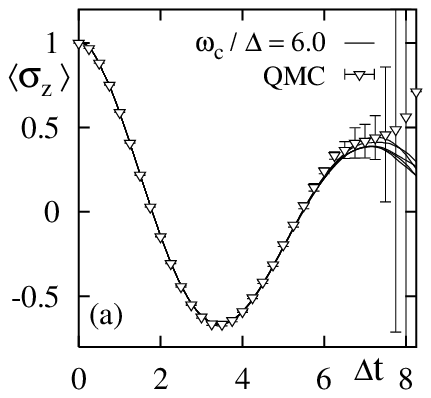}
\epsfxsize=0.5\columnwidth
\epsffile{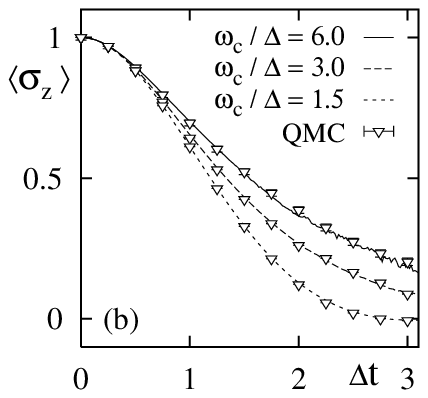}
}
\caption[]{Stochastic and quantum Monte Carlo simulations
  of a symmetric spin-boson system with $\chi_R(t)
  \propto t \exp(-\omega_c t)$, dissipation constant\cite{weiss99}
  $\alpha= 0.1$.
  (a) Coherent dynamics at zero
  temperature.
  (Five repeated simulations of Eq.\ (\ref{eq:linliou}) are shown to indicate
  statistical errors). (b) Incoherent dynamics at
$T=5\hbar\Delta/k_B$, simulated using Eq. (\ref{eq:eomfriction}).}
\label{fig:numerics}
\end{figure}

\end{multicols}
\vspace*{4mm}
\frenchspacing\widetext
{\raggedright \large\tt To be published in Phys. Rev. Lett. (2002)}
\end{document}